\documentclass{webofc}
\usepackage[varg]{txfonts}   
\def\s{{\,\rm s}}
\def\g{{\,\rm g}}
\def\eV{\,{\rm eV}}
\def\keV{\,{\rm keV}}

\def\GeV{\,{\rm GeV}}
\def\TeV{\,{\rm TeV}}
\def\sv{\left<\sigma v\right>}
\def\({\left(}
\def\){\right)}
\def\cm{{\,\rm cm}}

\def\beq{\begin{equation}}
\def\eeq{\end{equation}}
\begin{document}
\title{Cosmoparticle physics of dark matter}
%
%

\author{\firstname{Maxim} \lastname{Khlopov}\inst{1,3}\fnsep\thanks{\email{khlopov@apc.univ-paris7.fr}} 
}

\institute{Institute of Physics, Southern Federal University// Stachki 194 Rostov on Don 344090, Russia
\and
          National Research Nuclear University MEPhI (Moscow Engineering Physics Institute),// 115409
Moscow, Russia
\and
          APC laboratory 10, rue Alice Domon et Leonie Duquet// 75205 Paris Cedex 13, France
          }

\abstract{%
 The lack of confirmation for the existence of supersymmetric particles and Weakly Interacting Massive Particles (WIMPs) appeals to
extension of the field of studies of the physical nature of dark matter, involving non-supersymmetric and non-WIMP solutions.
We briefly discuss some examples of such candidates in their relationship with extension of particle symmetry and pattern of symmetry breaking. We specify in the example of axion-like particles nontrivial features of cosmological reflection of the structure and pattern of Peccei-Quinn-like symmetry breaking. The puzzles of direct and indiect dark matter searches can find solution 
in the approach of composite dark matter. The advantages and open problems of this approach are specified. We note that detailed analysis of cosmological consequences of any extension of particle model that provides candidates for dark matter inevitably leads to nonstandard features in the corresponding cosmological scenario. It makes possible to use methods of cosmoparticle physics to study physical nature of the dark matter in the combination of its physical, astrophysical and cosmological signatures.
}
\maketitle
\section{Introduction}
\label{intro}
We use the term "cosmoparticle physics" and not "astroparticle physics", which is widely used in English scientific literature. 
It is not a simple linguistic difference, the same as Russians say "cosmonauts", when English speaking people say "astronauts". Astroparticle physics is a very important part of cosmoparticle physics, but the latter involves much more general framework, involving cross disciplinary studies of fundamental relationship of cosmology and particle physics in the combination of its physical, astrophysical and cosmological probes \cite{ADS,MKH,book,newBook}. Methods of cosmoparticle physics are inevitable in studies of the physical nature of cosmological dark matter, dominating in the matter content of the modern Universe, as well as in general studies of physics beyond the Standard model (BSM) of elementary particles, on which modern cosmology is based. 

The data of precision cosmology \cite{planck15} prove existence of nonbaryonic dark matter that supports Cold Dark Matter (CDM) scenario of the Large Scale Structure (LSS) formation. During the last two decades the mainstream in the interpretation of Cold Dark Matter candidates was related with Weakly Interacting Massive Particles (WIMPs). It correlated with the hope to discover supersymmetric particles at the LHC. 

This approach was strongly motivated by the so called "WIMP miracle": the frozen out abundance of massive particles with masses of order tens-hundreds GeV, having annihilation cross section of the order of ordinary weak interaction, can naturally explain the measured dark matter density. Such particles were predicted in supersymmetric models as stable neutral lightest supersymmetric particles. Simultaneously it should have been accompanied by a set of supersymmetric partners of ordinary particles to be discovered in the experiments at the LHC.

It is the advantage of supersymmetric (SUSY) models with the scale of SUSY breaking in the range of hundreds GeV to solve the internal problem of the Standard model, related with the divergence of mass of Higgs boson. Due to the opposite sign of contributions by SUSY partners divergent loop diagrams with ordinary particles are canceled in the radiative contributions to the Higgs boson mass. Renormalization group arguments can also explain the origin of the Higgs form of potential of scalar field and thus the physical nature of the electroweak symmetry breaking.

Neither SUSY particles are still found at the LHC, nor WIMPs are definitely detected in the underground direct searches. It implies extension of the field of studies of possible dark matter candidates, strongly motivated by particle theory, involving their non-SUSY nature and non-WIMP features.

Here we give following \cite{khlsym,me} a brief overview of theoretical motivation for such candidates (Sect.~\ref{sec-1}) and then discuss in more details nontrivial features of axion-like (Sect.~\ref{sec-2}) and composite dark matter models (Sect.~\ref{sec-3}). We come to conclusion in Sect.~\ref{sec-4} that combination of physical astrophysical and cosmological probes will inevitably shed light on the new physics, on which modern cosmology is based.
\section{Extending the field of dark matter physics}
\label{sec-1}
Dark matter candidates should be sufficiently long-living to survive to the present time or at least to the period of cosmological large scale structure formation, since they should trigger development of gravitational instability before recombination of hydrogen. This role implies dark matter decoupling from plasma and radiation before the beginning of matter dominated stage.

The condition of stability can be satisfied in a wide range of extensions of the Standard model, provided that they involve new strict or nearly strict symmetry that acts on  new set of particles. Then the lightest particle that possess such symmetry is absolutely stable, if the symmetry is strict, or sufficiently long living, if the conservation of the new charge, related with the new symmetry, can be violated. 

The condition of decoupling implies {\it sufficiently weak} interaction of dark matter candidates with plasma and radiation. In the early Universe the term "sufficiently weak" corresponds to rather wide range of cross sections for dark matter particle interaction with ordinary matter: from a super weak to the nuclear strong interaction. Even ordinary hadronic cross sections are sufficiently weak for dark matter decoupling from the low number density baryonic plasma \cite{hadronic} (see also \cite{beylin,kuksa} and conributions \cite{beylin1,kuksa1} to these proceedings). 

\subsection{Massive neutrino}
\label{nu}
Massive neutrinos were historically first candidates for dark matter (see e.g. \cite{book,newBook} for review and reference). Big Bang theory gives very definite prediction for the number density $n_{\nu \bar{\nu}}$ of primordial known active neutrino species: 
\beq n_{\nu \bar{\nu}}= \frac{3}{11}n_{\gamma},\label{nua}\eeq where $n_{\gamma}$ is the number density of the CMB photons. Multiplied by the value of neutrino mass the Eq. \ref{nua} defines the cosmological density of massive neutrino.

Direct experimental upper limit on the mass of electron neutrino together with the data on neutrino oscillations put upper limit on this density, which is much below the measured dark matter density.

However, though the nonzero mass of neutrino is proved by neutrino oscillations, the physical nature of neutrino mass is still unknown.
The mechanism of neutrino mass generation may involve sterile (right-handed) neutrino, which can explain all the measured dark matter density \cite{nus}. Therefore massive known neutrinos cannot be viable dark matter candidates, but the BSM physics of neutrino mass can provide the solution for the dark matter problem.
It is interesting to note that the dark matter scenario with sterile neutrino inevitably contains some features, going beyond the standard cosmological model \cite{nus}.
\subsection{SUSY at subplanckean scale?}
\label{sugra}
The lack of positive evidence for SUSY at the LHC energy moves the SUSY scale to higher energies. In the extreme case it may be as high as the scale of Grand Unified Theories (GUT) and linked to unification of all the four fundamental forces, including gravity in the framework of Supergravity
(see \cite{me} and \cite{ketovsym} for review and references). 

Such models can predict stable superheavy gravitino as dark matter candidate \cite{ketov} as well as provide physical basis for Starobinsky supergravity \cite{ketov2}, reproducing Starobinsky inflation scenario \cite{star} on supergravity basis. Probes for such models are related to their cosmological consequences that can lead to observable effects, like Primordial Black holes and effects of their evaporation or superheavy dark matter effect in ultra high energy cosmic rays \cite{ketovsym,me,PBHrev}.

In the minimal Starobinsky-Polonyi N=1 Supergravity superheavy gravitino are produced in decays of inflaton and Polonyi fields. It links physics of inflation to physics of superheavy gravitino dark matter in these models (see \cite{ketov} for review and references).

However, the advantage of solution for the problem of diveregence of the Higgs boson mass and the origin of electroweak symmetry breaking is lost in the SUSY models with super high energy scale. It appeals to non-SUSY solution for this problem.
\subsection{Composite Higgs}
\label{higgs}
In all the approaches to solve the problem of the Higgs boson and origin of electroweak symmetry breaking \cite{kaplan1,kaplan2,arkani1,arkani2,pomarol1,pomarol2} new particles are predicted. In particular, the approach \cite{CTTP}, involving colored twisted partners of top quark, can lead to existence of colored particles with arbitrary charge.
In this approach colored particles with charge that differs from $Q=2/3+n$ or from $Q=-(n+1/3)$ cannot decay to SM particles and should be stable or long living \cite{partnerium}.

The problem of divergence of the mass of an elementary Higgs boson, can find solution, if Higgs boson is composite. Then the electroweak symmetry breaking scale and the mass of Higgs boson are determined by the scale, at which the constituents are confined. The origin of this scale may come from confinement of new nonabelean gauge charges, similar to the QCD case. This idea of a 'technicolor' is now developed in the Walking Technicolor models, in which gauge constants are not running, but walking \cite{Sannino:2004qp,Hong:2004td,Dietrich:2005jn,Dietrich:2005wk,Gudnason:2006ug,Gudnason:2006yj}. This nontrivial dependence of the gauge coupling on distance makes possible to arrange the scale $\Lambda_{WTC}$ of technicolor confinement above electroweak scale $v$ even if technicolor symmetry is based on $SU(2)$ gauge group. Note that $SU(2)$ confinement with a running constant would take place at macroscopical distances and that would be the case for electroweak $SU(2)$ symmetry without Higgs mechanism.

Together with binding in the composite Higgs boson, technicolor constituents can also form a set of new composite particles. At the energy $E\ll \Lambda_{WTC}$ their composite nature is not feasible and they look like new elementary particles. Since techniquarks have only technicolor and do not possess ordinary QCD color, their bound states look like ordinary leptons.
\subsection{Mirror world?}
\label{mirror}
Extensive hidden sector of particle theory can provide the existence of new interactions, which only new particles possess. It gives rise to Self Interacting Dark Matter as well as to Composite Dark Matter, in which dark matter species are composed of constituents, bound by various forces. 

Historically one of the first examples of such composite dark matter was presented by the model of mirror matter. Mirror particles, first proposed by T. D. Lee and C. N. Yang in \cite{LeeYang} to restore equivalence of left- and right-handed co-ordinate systems in the presence of P- and C- violation in weak interactions, should be strictly symmetric by their properties to their ordinary twins. After discovery of CP-violation it became clear that antiparticles cannot play the role of mirror partners. It was shown by I. Yu. Kobzarev, L. B. Okun and I. Ya. Pomeranchuk in \cite{KOP} that mirror partners should represent a new set of symmetric partners for ordinary quarks and leptons with their own strong and electromagnetic mirror interactions. Discovery of weak Z boson and measurement of its properties excluded common weak interaction, so that mirror partners should have their own mirror weak interaction. It means that there should exist mirror quarks, bound in mirror nucleons by mirror QCD forces and mirror atoms, in which mirror nuclei are bound with mirror electrons by mirror electromagnetic interaction \cite{ZKrev,FootVolkas}. If gravity is the only common interaction for ordinary and mirror particles, mirror matter can be present in the Universe in the form of elusive mirror objects, having symmetric properties with ordinary astronomical objects (gas, plasma, stars, planets...), but causing only gravitational effects on the ordinary matter\cite{Blin1,Blin2}.

Even in the absence of any other common interaction except for gravity, the observational data on primordial helium abundance and upper limits on the local dark matter seem to exclude mirror matter, evolving in the Universe in a fully symmetric way in parallel with the ordinary baryonic matter\cite{Carlson,FootVolkasBBN}. The symmetry in cosmological evolution of mirror matter can be broken either by initial conditions\cite{zurabCV,zurab}, or by breaking mirror symmetry in the sets of particles and their interactions as it takes place in the shadow world \cite{shadow,shadow2}, arising in the heterotic string model. We refer to Refs.
\cite{newBook,OkunRev,Paolo} for review of mirror matter and its cosmological probes.

For the general case of the shadow world the detailed analysis becomes rather complicated, involving general qualitative astrophysical constraints \cite{shadow,shadow2}. Some examples of dark atoms bound by various forces are discussed in Ref. \cite{adm}. 

\subsection{Primordial black holes}
\label{pbh}
In the modern Universe black holes with the mass above few Solar mass are expected to be formed in the final stage of evolution of massive stars \cite{1,ZNRA}. However, in the early Universe primordial black holes (PBH) with much smaller mass can be formed, if cosmological expansion has stoped in some region within the cosmological horizon \cite{ZN,hawking1,hawkingCarr}. It corresponds to very strong nonhomogeneity of expansion and in the homogeneous and isotropic Universe probability of PBH formation is determined by the equation of state.

In the universe with equation of state \begin{equation}
p=\gamma \epsilon \label{EqState} \end{equation} with numerical
factor $\gamma$ being in the range \begin{equation}
\label{FacState}0 \le \gamma \le 1\end{equation} the probability of
forming a black hole from fluctuations within the cosmological horizon is
given by \cite{carr75}
\begin{equation}
W_{PBH} \propto \exp \left(-\frac{\gamma^2}{2
\left\langle \delta^2 \right\rangle}\right),
\label{ProbBH}
\end{equation}
where $\left\langle \delta^2 \right\rangle \ll 1$ is the amplitude of density fluctuations. In the hot Unvierse with relativistic equation of state ($\gamma = 1/3$) the probability (\ref{ProbBH}) is exponentially small and can be significant only provided that the spectrum of density fluctuations is not flat and has a strong enhancement at small scales.

Cosmological consequences of BSM physics can lead to nontrivial scenarios of very early Universe in which various mechanisms of primordial black hole  formation are possible. It makes the constraints on the PBH spectrum an effective probe for BSM physics at superhigh energy \cite{PBHrev}. 

Early matter dominance, corresponding to $\gamma=0$ in the Eq.(\ref{ProbBH}) removes exponential suppression in the probability of PBH formation and triggers development of gravitational instability, in which stromg inhomogeneities can grow from small initial fluctuations. Spectrum of PBH masses is sensitive to the properties of particles or fields, dominating at this stage \cite{PBHrev,khlopov0,khlopov1,Kalashnikov,Kadnikov}.

The pattern of symmetry breaking leads to a succession of phase transitions. If phase transitions are of the strong first order, bubble nucleation can lead to copious PBH formation in bubble collisions \cite{kkrs}.

PBHs with mass $M \le 10^{15} \g$ do not survive to the present time evaporating by the mechanism of Hawking \cite{hawking4,21}. This process is the universal process of production of any type of particles with mass 
$$m \le T_{evap} \approx 10^{13} \GeV\frac{1\g}{M}$$
that are predicted in our space-time and can be the source of any superweakly interacting particles, like gravitino \cite{KBgrain}.

PBHs with mass $M \ge 10^{15} \g$ should survive to the present time and can play the role of dark matter. 

\section{Primordial structures from axion-like models}
\label{sec-2}
\subsection{Pseudo Nambu-Goldstone field}
\label{pngb}
Axion-like models can be illustrated by a simple model of a complex field $\Psi = \psi \exp{i \theta}$ with broken global $U(1)$ symmetry \cite{book2,dmrps}. The potential
$$V=V_0+\delta V$$
contains the term 
\beq V_0= \frac{\lambda}{2}(\Psi^{\ast} \Psi -f^2)^2 \label{vo} \eeq
that leads to spontaneous breaking of the $U(1)$ symmetry with continuous degeneracy of the asymmetric ground state
\beq \Psi_{vac}= f \exp (i \theta) \label{cav} \eeq
and the term 
\beq \delta V (\theta)= \Lambda^4 (1-\cos\theta) \label{disc} \eeq\
with $\Lambda \ll f$ that leads to manifest breaking of the residual symmetry, leading to a discrete set of degenerated ground states, corresponding to 
$$\theta_{vac} = 0, 2\pi, 4\pi, ...$$
The term (\ref{disc}) can be present in the theory initially, or generated by instanton transitions, as it is the case in the axion models.
In the result of the second step of symmetry breaking a pseudo-Nambu-Goldstone field $\phi = f \theta$ is generated with the mass \beq m_{\phi} = \Lambda^2/f. \label{pngm} \eeq The existence of $\phi \gamma \gamma$ vertex leads to a two-photon decay of $\phi$, as well as to effects of $\phi \gamma$ conversion \cite{raffelt} like axion-photon conversion in magnetic field (see e.g. \cite{soda} for review and references). The principles of experimental search for axion by "light shining through walls" effects are based on such a conversion \cite{sikivie}.

In axion models the mass of  axion is given by \beq
m_a=C m_{\pi}f_{\pi}/f,\label{axion}\eeq where $m_{\pi}$ and $f_{\pi}\approx m_{\pi}$ are the pion mass and constant, respectively, the constant $C\sim 1$ depends on the choice of the axion model and $f\gg f_{\pi}$ is the scale of the Peccei-Quinn symmetry breaking). The relationship (\ref{axion}) of axion to neutral pion makes possible to estimate the cross section of axion interactions from the corresponding cross section of pion processes multiplied by the factor $(f_{\pi}/f_a)^2$.

Axion couplings to quarks and leptons can lead to rare processes like $K \rightarrow \pi a$ or $\mu \rightarrow e a$. In the gauge model of family symmetry breaking \cite{berezhiani5} the PNG particle called \textit{archion}  shares properties of axion with the ones of singlet Majoron and familon, being related to the mechanism of neutrino mass generation. In this model together with archion decays of quarks and charged leptons archion decays $\nu_H \rightarrow \nu_L a$ of heavier neutrino $\nu_H$ to lighter neutrino $\nu_L$ are also predicted \cite{berezhiani5}.

\subsection{Archioles}
\label{archioles}
If the reheating temperature $T_{reh}$ exceeds $f$
the $U(1)$ symmetry breaking pattern implies the
succession of second order phase transitions. 

In the first
transition, continuous degeneracy of vacua leads---at scales
exceeding the correlation length---to the formation of topological
defects in the form of a string network; in the second phase
transition, continuous transitions in space between degenerated
vacua form surfaces: domain walls surrounded by strings. This last
structure is unstable, but, as was shown in the example of the
invisible axion \cite{Sakharov2,kss,kss2}, it is reflected in the
large scale inhomogeneity of the distribution of the energy density of
coherent PNG (axion) field oscillations. This energy density is
proportional to the initial phase value, which acquires a dynamical
meaning of the amplitude of the axion field, when axion mass is switched on
as a result of the second phase transition. In spite of a very small mass of PNG particles, geven by Eq.(\ref{pngm}) $m_{\phi}=\Lambda^2/f$  with $f \gg \Lambda$, these particles are created in Bose-Einstein condensate in the ground state, i.e. they are initially created as nonrelativistic in the very early Universe. This feature, typical for invisible axion models, can be the general feature for all the axion-like PNG particles.

The value of phase changes by $2 \pi$ around string. This strong
nonhomogeneity of phase leads to corresponding nonhomogeneity of
the energy density of coherent PNG (axion) field oscillations. The usual
~argument (see \cite{kim} and references therein) is that this strong inhomogeneity is essential
only at scales corresponding to the mean distance between strings.
This distance is small, being of the order of the scale of
the cosmological horizon in the period when PNG field oscillations
start. However, since the nonhomogeneity of phase follows the
pattern of the axion string network, this argument misses large scale
correlations in the distribution of oscillations' energy density.

Indeed, numerical analysis of the string network (see the review in \cite{vs}) indicates that large string loops are strongly suppressed,
and the fraction of about 80\% of string length (corresponding to
long loops) remains virtually the same in all large scales. This
property is the other side of the well known scale invariant
character of the string network. Therefore, the correlations of energy
density should persist on large scales, as was revealed in Refs.
\cite{Sakharov2,kss,kss2}. Evolution of such correlations and their relationship with axion or boson star formation \cite{tkacheva,tkachevb}
deserves attention and special studies.

If phase transitions take place after the reheating of the Universe, large
scale correlations in topological defects and their imprints in
primordial inhomogeneities are the indirect effect of inflation.
Inflation provides, in this case, the equal conditions of phase
transition, taking place in causally disconnected~regions.

\subsection{Clusters of PBHs}
\label{cpbh}
\begin{figure}[htbp]
\centering
\includegraphics[width=12cm,clip]{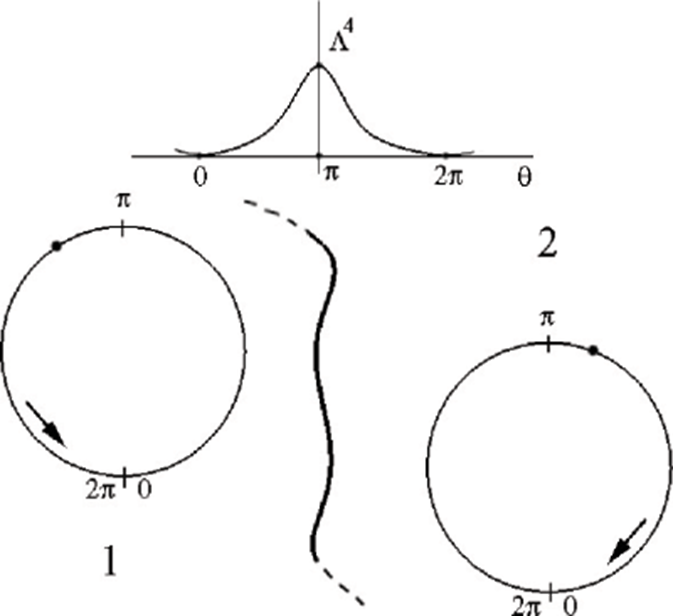}
\caption{The inflational  evolution of the
phase (taken from Reference \cite{book2}), resulting after crossing $\pi$ in appearance of region 1 with $\theta_{vac}=2\pi$, surrounded by space 2 with $\theta_{vac}=0$. After the second phase transition at the border of the region 1 a closed wall is formed. }
\label{fig-1}       
\end{figure}

If the first phase transition takes place at the inflationary stage with Hubble parameter $H_{infl}$, the phase value $\theta_{60}$, corresponding to the $e$-folding, at which the now observed part of the Universe was formed, fluctuates at successive steps of inflation. 
Therefore if the phase value $\theta_{60}$ is in the range $[0 ,\pi ]$, it makes Brownian steps
$\delta\theta_{e}=H_{infl}/(2\pi f)$  at each succesive e--folding, corresonding to smaller parts of the now observed Universe. The
typical wavelength of the fluctuation $\delta\theta$ is equal to
$H^{-1}_{infl}$. The whole domain  $H^{-1}_{infl}$ containing phase
$\theta_{N}$ gets  divided (after the next e-folding) into  $e^3$ causally
disconnected domains of radius $H^{-1}_{infl}$. Each new domain
contains almost homogeneous phase value
$\theta_{N-1}=\theta_{N}\pm\delta\theta_{e}$. This process, illustrated by Fig. \ref{fig-1}, repeats in every domain with every successive e-folding.
and can cross $\pi$, so that regions with true vacua $0$ and $2\pi$, separated by closed walls, appear in the result of the second phase transition. After the first crossing of $\pi$ successive crossings can take place at the next steps of inflation, leading to smaller closed walls, clustering around the biggest one.
Collapse of these close walls leads to formation of clusters of black holes with the mass, which is determined by PNG parameters $\Lambda$ and $f$ \cite{book2,dmrps,pbhClusters}.

The maximal BH mass is
determined by the condition that the wall does not dominate locally
before it enters the cosmological horizon. Otherwise, local wall
dominance leads to a superluminal $a \propto t^2$ expansion for the
corresponding region, separating it from the other part of the
universe. This~condition corresponds to the mass \cite{book2}\beq
\label{Mmax} M_{max} =
\frac{m_{pl}}{f}m_{pl}(\frac{m_{pl}}{\Lambda})^2\eeq 

The minimal
mass follows from the condition that the gravitational radius of BH
exceeds the width of wall, and it is equal to \cite{book2}\beq
\label{Mmin} M_{min} = f(\frac{m_{pl}}{\Lambda})^2\eeq

Closed wall collapse leads to primordial gravitational wave (GW) 
 spectrum, peaked at \beq
\label{nupeak}\nu_0=3 \times 10^{11}(\Lambda/f)\,{\rm Hz} \eeq with
energy density up to \beq \label{OmGW}\Omega_{GW} \approx
10^{-4}(f/m_{pl})\eeq 

At $f \sim 10^{14}$ GeV this primordial
gravitational wave background can reach $\Omega_{GW}\approx
10^{-9}.$ For the physically reasonable values of \beq
1<\Lambda<10^8\,{\rm GeV}\eeq the maximum of the spectrum corresponds to
\beq 3 \times 10^{-3}<\nu_0<3 \times  10^{5}\,{\rm Hz}\eeq 

In the range from tens to 
thousands of Hz, such background may be a challenge for  LIGO 
 experiment. 
 
Critical anlysis by \cite{pbhClusters} of constraints \cite{CarrKeun} on relative contribution of PBHs into dark matter density, excluding PBH dominance in dark matter, shows that PBH formation in clusters can influence the constraints \cite{CarrKeun} and even open a possiblity for PBH dominant dark matter.

 Another
profound signature of the considered scenario are gravitational wave
signals from merging of BHs in PBH cluster. Being in cluster, PBHs with the masses of tens $M_{\odot}$ form binaries much easier, than in the case of their random distribution. In this aspect detection of signals from binary BH coalescence in the gravitational wave experiments \cite{gw1,gw2,gw3,gw4,gw5} may be considered as a positive evidence for this scenario. Repeatedly detected signals localized in the same place would provide successive support in its favor or exclusion \cite{pbhClusters,Bringmann}. The~existing statistics is evidently not sufficient to make any definite conclusion on this possibility. However, repeating detection of four GW signals in the August of 2017 noted in GWTC catalog \cite{GWTC} may be an interesting hint to such a possibility \cite{me}. 

\section{Dark atom solution for dark matter puzzles}
\label{sec-3}
\subsection{Dark atoms}
Atoms of dark matter, in which new stable charged particles are bound by Coulomb force, were first proposed in \cite{Glashow:2005jy}. However, it was shown in \cite{BKSR1} that prediction of single charged stable particles inevitably leads to overproduction of anomalous hydrogen, severely constrained in the terrestrial matter. Stable particles with charge $+1$ bind with electrons, forming atoms of anomalous hydrogen. Stable particles with charge $-1$ bind with primordial helium in $+1$ charged ion  immediately after Big Bang Nucleosynthesis, and this $+1$ charged ion also forms with electron anomalous hydrogen. Only stable particles with even negative charge $-2n$ avoid this immediate trouble, bound with primordial helium in neutral atoms \cite{I}. 

However, existence of particles with charge $-2,-4...$ should be inevitably accompanied by the existence of the corresponding positively charged particles that can bind with electrons and form anomalous helium, berillium etc, respectively. The abundance of such anomalous isotopes can be suppressed in the terrestrial matter, if new particles have additional long range interaction making positively charged constituents of anomalous isotopes to recombine with the corresponding negatively charged partilces in the terrestrial matter, thus reducing their abundance below the experimental upper limits \cite{FKS}. 

In WTC $-2n$ charged stable techniparticles can be generated in excess over their $+2n$ charged partners, equilibrated by sphaleron transitions with the baryon excess. The relationship between the excess of $-2n$ and baryon asymmetry can explain the observed ratio of baryonic and dark matter densities \cite{me,KK,KK1}. In this case dark matter is in the form of dark atoms, in which the excessive $-2n$ charged particles are bound by Coulomb force with $n$ helium nuclei. Such systems look like Thomson atom: $-2n$ charged leptom is inside of nuclear droplet formed by $n$ nuclei of helium.
Studies of composite dark matter involving multiple charged particles is now under way, but we can illustrate the main features of dark atom scenario by the example of $O$He model, in which bound states of $-2$ charged particles $O^{-2}$ and primordial helium are proposed as the candidates for the dominant dark matter component.

In the case of $O$He Bohr-atom-like description is appropriate \cite{khlsym,I,KK,pos}.
The radius of Bohr orbit in these ``atoms''
\cite{I,KK,pos} is \beq
r_{o} = \frac{1}{Z_{o} Z_{He} \alpha 4 m_p} = 2 \cdot 10^{-13}\cm, 
\eeq
being of the order of and even a bit smaller than the size of He nucleus. Therefore non-point-like charge distribution in He leads to a significant correction to the \textit{O}He binding energy.

 In contrast to the ordinary atoms, having electroweakly interacting shell and the core much smaller, than the atomic size, \textit{O}He has strongly interacting helium shell and the size of the orbiting He is of the order of radius of orbit. Therefore, in the lack of these usual approximations of atomic physics proper description of \textit{O}He interaction with nuclei remains an open problem.
 The most complicated problem of $O$He-nucleus interaction is the self consistent treatment of simultaneous action of nuclear attraction to the He shell of approaching nucleus and its Coulomb repulsion. 

\subsection{Dark atom cosmology}
\label{dac}
$O$He cosmological evolution  \cite{khlsym,I,FKS,KK,KK1,pos,Khlopov:2010ik,Khlopov:2011me,mpla,khlopov.proc.2014,khlopov.proc.2015,khlopov.ijmpd.2015,probes}
starts by creation of \textit{O}He as soon as primordial helium is produced in Big Bang Nucleosynthesis. Since the mass of $O^{--}$ is in TeV range, the number density of these particles is by two orders of magnitude smaller, than He abundance, and the frozen out concentration of free $O^{--}$ is exponentially suppressed. In principle, $O$He interaction with nuclei can catalize formation of primordial heavy elements, but the corresponding analysis implies further developement of $O$He nuclear physics.

Due to elastic nuclear interactions of helium shell with nuclei (dominantly protons) \textit{O}He gas is in thermal equilibrium with plasma and radiation, while energy and momentum transfer from plasma is effective. In this period radiation pressure is transferred to \textit{O}He
density fluctuations, transforming them in acoustic waves at scales up to the size of the cosmological horizon.

The energy and momentum transfer from baryonic plasma to \textit{O}He cannot support thermal equilibrium at temperature $T < T_{od} \approx 1 S_3^{2/3}\keV$ 
\cite{I,KK} since \beq n_B \sv (m_p/m_o) t < 1, \eeq where $m_o$ is the
mass of $O^{--}$, which determines the mass of $O$He atom and $S_3= m_o/(1 \TeV)$. Here $n_B$ is the baryon number density, the cross section is given by \beq \sigma
\approx \sigma_{o} \sim \pi r_{o}^2 \approx
10^{-25}\cm^2\label{sigOHe}, \eeq and $v = \sqrt{2T/m_p}$ is the
baryon thermal velocity. Then $O$He gas decouples from plasma and
starts to dominate in the Universe after $t \sim 10^{12}\s$  at $T
\le T_{RM} \approx 1 \eV$.

Decoupled from plasma and radiation $O$He atoms play the role of dark matter in the development of gravitational instability, triggering large scale structure formation. Nuclear interacting nature of $O$He determines specifics of its spectrum of density fluctuations. Conversion in sound waves leads to suppression on the corresponding scales and the spectrum acquires the features of Warmer than Cold Dark Matter scenario \cite{khlsym,I,FKS,KK,pos}. Decoupled from baryonic matter $O$He gas doesn't follow formation of baryonic objects, forming dark matter halos of galaxies.

In spite of strong (hadronic) cross section $O$He gas is collisionless on the scale of galaxies, since its collision timescale is much larger than the age of the Universe. Baryonic matter in the Galaxy is also tranparent in the average, so that $O$He can be captured only by sufficiently dense matter proto-object clouds and objects, like planets and stars. 
\subsection{Dark atoms in underground detectors}
\label{dau}
In terrestrial matter such dark matter species are slowed down and cannot cause significant nuclear recoil in the underground detectors, making them elusive in direct WIMP search experiments (where detection is based on nuclear recoil) such as CDMS, XENON100 and LUX. The positive results of DAMA/NaI and DAMA/LIBRA experiments (see \cite{DAMAtalk} for review and references) can find in this scenario a nontrivial explanation due to a low energy radiative capture of $OHe$ by intermediate mass nuclei~\cite{mpla,DMRev,DDMRev}. This explains the negative results of the XENON100 and LUX experiments with heavy element content. The rate of this capture is proportional to the temperature: this leads to a suppression of this effect in cryogenic detectors, such as CDMS.
The main problem of this scenario is the lack of correct quantum-mechanical treatment of the complicated problem of $O$He-nuclear interaction.
\subsection{Indirect effects of dark atoms}
\label{ida}
Being asymmetric dark matter, $O$He collisions cannot lead to indirect effects like WIMP annihilation  (first considered in \cite{ZKKC}) contributing by its products to gamma background and cosmic rays. However, $O$He excitations in such collisions can result in pair production in the course of de-excitation and the estimated emission in positron annihilation line can explain the excess, observed by INTEGRAL in the galactic bulge \cite{CKWahe}. For realistic estimation of the dark matter density in the center of Galaxy such explanation is possible for a narrow range of $O^{--}$ mass near 1.25 TeV \cite{me,front}.

In the two-component dark atom model, based on the Walking Technicolor, together with the dominant component of $O$He a subdominant WIMP-like component $UU\zeta$ is present, with metastable technibaryon $UU$, having charge +2. Decays of this technibaryon to the same sign (positive) lepton pairs
can explain excess of high energy cosmic positrons observed by PAMELA and AMS02 \cite{AHEP}. However, any source of positrons inevitably is also the source of gamma radiation. Therefore the observed level of gamma background puts upper limit on the mass of $UU$, not exceeding 1 TeV \cite{front}.

The problem of gamma radiation overproduction for dark matter interpretation of high energy positron excess seems to bear general character for any extensive distribution of the dark matter (see Ref. \cite{kostya} for recent review). In the essence it follows from the fact that due to energy losses and diffusion in galactic magnetic fields high energy positrons come from the sources situated in the neighbourhood of the Solar system, while all such sources distributed over the Galaxy contribute to the gamma ray background. Even clumpiness of such distribution can hardly help: one needs strong enhancement of positron production in the local clumps with simultaneously strong suppression of gamma ray emission from distant clumps in the Galaxy.
\subsection{Search for multiple charged dark atom constituents at the LHC}
\label{lhc}
The widely accepted approach to collider search for dark matter particles is related to search for effects of missing energy and momentum in particle collisions. This approach links freezing out of dark matter partilces in early Universe, indirect effects of their annihilation in the Galaxy  and direct underground searches for dark matter particles. 

\begin{figure}[htbp]
\centering
\includegraphics[width=12cm,clip]{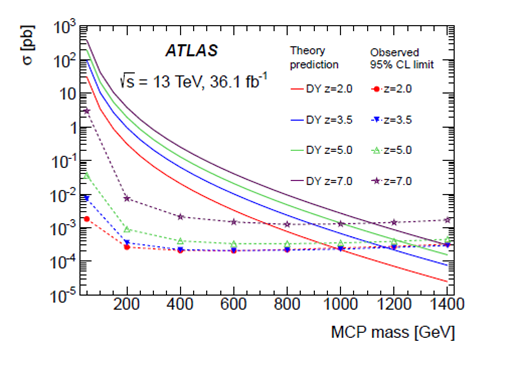}
\caption{Search for heavy long-lived multi-charged particles in proton-proton collisions at $\sqrt{s} = 13~\TeV$ using the ATLAS detector \cite{Aad2019}. Observed $95\%$ CL cross-section upper limits and theoretical cross-sections as functions of the multi-charged particles mass. The~double charged particles are denoted as ``$z=2$'' (red points and lines). The~picture is taken following \cite{me} from~Ref.\cite{Aad2019}.}
\label{fig-2}       
\end{figure}

Such relationships are strongly modified in the case of dark atoms. In particular, in the context of dark atom model, accelerator probe for dark matter is reduced to search for  stable double charged constituents of composite dark matter that acquires the significance of direct experimental test for dark atom model. Upper limits on the mass of double charged particles, at which the detected anomalies in positron annihlation line radiation from the center of Galaxy ($m \approx 1.25 \TeV$) and the excess of cosmic high energy positrons ($m<1 \TeV$) find explanation in the framework of dark atom model provide a possibility of \textit{experimentum crucis } for such explanation \cite{2cBled}. The recent lower mass limit, presented by ATLAS collaboration in \cite{Aad2019}, is set at $980$GeV at $95\%$ CL. The~comparison of observed cross-section upper limits and theoretically predicted cross-sections is shown in Fig.~\ref{fig-2}. These results exclude explanation of excess of cosmic high energy positrons by decay of double charged dark atom constituents with the mass below $1 \TeV$ and approach complete test for composite dark matter nature of anomaly in positron annihlation line radiation from the center of Galaxy.

\section{Conclusion}
\label{sec-4}
The existence of nonbaryonic dark matter is the cornerstone of the modern standard cosmology. It implies the existence of physics beyond the Standard model of elementary particles. However, the results of the experiments at the LHC put only more and more stringent constraints on the effects of BSM physics. One can call it {\it conspiracy of BSM physics} \cite{me}. 

On the other hand, there is {\it conspiracy of BSM cosmology}. Precision cosmology only strengthens the constraints on possible nonstandard cosmological scenarios. 

The present brief review shows that any physically motivated extension of the particle Standard model predicts together with dark matter candidates a set of model dependent predictions of new physics effects, involving, in particular, observable nonstandard features in cosmological scenario.

It inspires the hope that methods of cosmoparticle physics will remove the conspiracy of BSM physics of dark matter and reveal its relationship with the BSM cosmology in the combination of physical, astrophysical and cosmological probes.
\section*{Acknowledgements}
The work was supported by grant of Russian Science Foundation (Project No-18-12-00213).

%
%

\end{document}